\documentclass[conference, final]{IEEEtran}
\IEEEoverridecommandlockouts
\usepackage{url}
\usepackage{booktabs}
\usepackage{amsmath,amssymb,amsfonts}
\usepackage{xcolor}
\usepackage{tikz}
\usepackage{array}
\usepackage{caption, subcaption}
\usepackage[noadjust]{cite}

\newcommand\anon[2]{#2} % #1==ANON; #2==WITH NAMES
\newcommand\haveSAT[2]{#2} % #1==ANF only; #2==ANF+CNF
\newcommand{\Bosphorus}{\textsc{Bosphorus}}

\newcommand\cut[1]{}
\newcolumntype{H}{>{\setbox0=\hbox\bgroup}c<{\egroup}@{}}

\usetikzlibrary{positioning,shapes,arrows,fit,matrix,calc,decorations.pathreplacing,shapes.symbols,shapes.geometric}
\tikzset{>=latex}
\tikzstyle{decision} = [diamond, draw, align=center]
\tikzstyle{block} = [rectangle, draw, align=center, minimum height=3em, minimum width=6em, text centered]
\tikzstyle{cloud} = [draw, ellipse]
\tikzstyle{document} = [shape=tape, draw, align=center, minimum height=3em, minimum width=6em, , tape bend top=none]
\tikzstyle{input} = [shape=trapezium, draw, align=center, , minimum height=3em, minimum width=6em, trapezium left angle=60, trapezium right angle=120]

\newenvironment{Karnaugh}%
{
	\begin{tikzpicture}[scale=0.8, baseline=(current bounding box.north)]
	\draw (0,0) grid (4,4);
	\matrix (mapa) [matrix of nodes,
	column sep={0.8cm,between origins},
	row sep={0.8cm,between origins},
	every node/.style={minimum size=0.3mm},
	anchor=8.center,
	ampersand replacement=\&] at (0.4,0.4)
	{
		\& |(c00)| $\bar{x}_3 \bar{x}_4$         \& |(c01)| $\bar{x}_3 x_4$         \& |(c11)| $x_3 x_4$         \& |(c10)| $x_3 \bar{x}_4$         \& |(cf)| \phantom{00} \\
		|(r00)| $\bar{x}_1 \bar{x}_2$             \& |(0)|  \phantom{0} \& |(1)|  \phantom{0} \& |(3)|  \phantom{0} \& |(2)|  \phantom{0} \&                     \\
		|(r01)| $\bar{x}_1 x_2$             \& |(4)|  \phantom{0} \& |(5)|  \phantom{0} \& |(7)|  \phantom{0} \& |(6)|  \phantom{0} \&                     \\
		|(r11)| $x_1 x_2$             \& |(12)| \phantom{0} \& |(13)| \phantom{0} \& |(15)| \phantom{0} \& |(14)| \phantom{0} \&                     \\
		|(r10)| $x_1 \bar{x}_2$             \& |(8)|  \phantom{0} \& |(9)|  \phantom{0} \& |(11)| \phantom{0} \& |(10)| \phantom{0} \&                     \\
		|(rf) | \phantom{00}   \&                    \&                    \&                    \&                    \&                     \\    
	};
}%
{
	\end{tikzpicture}
}

\newcommand{\contingut}[1]{%
	\foreach \x [count=\xi from 0]  in {#1}
	\path (\xi) node {\x};
}

\newcommand{\implicant}[3][0]{
	\draw[rounded corners=3pt] ($(#2.north west)+(135:#1)$) rectangle ($(#3.south east)+(-45:#1)$);
}

\newcommand{\implicantcostats}[3][0]{
	\draw[rounded corners=3pt] ($(rf.east |- #2.north)+(90:#1)$)-| ($(#2.east)+(0:#1)$) |- ($(rf.east |- #3.south)+(-90:#1)$);
	\draw[rounded corners=3pt] ($(cf.west |- #2.north)+(90:#1)$) -| ($(#3.west)+(180:#1)$) |- ($(cf.west |- #3.south)+(-90:#1)$);
}

\newcommand{\implicantdaltbaix}[3][0]{
	\draw[rounded corners=3pt] ($(cf.south -| #2.west)+(180:#1)$) |- ($(#2.south)+(-90:#1)$) -| ($(cf.south -| #3.east)+(0:#1)$);
	\draw[rounded corners=3pt] ($(rf.north -| #2.west)+(180:#1)$) |- ($(#3.north)+(90:#1)$) -| ($(rf.north -| #3.east)+(0:#1)$);
}
\usepackage{authblk}

\title{%
{\Bosphorus}: Bridging ANF and CNF Solvers\thanks{The open-source tool is available at \protect\url{https://github.com/meelgroup/bosphorus}}
}
\author[*]{Davin Choo}
\author[$\dagger$]{Mate Soos}
\author[*]{Kian Ming A. Chai}
\author[$\dagger$]{Kuldeep S. Meel}
\affil[*]{\textit{Information Division},
	\textit{DSO National Laboratories, Singapore}}
\affil[$\dagger$]{	\textit{School of Computing},
	\textit{National University of Singapore}}

\begin{document}
	\maketitle
	
	\begin{abstract}
		Algebraic Normal Form (ANF) and Conjunctive Normal Form (CNF) are commonly used to encode problems in Boolean algebra.
		ANFs are typically solved via Gr\"obner basis algorithms, often using more memory than is feasible;
		while CNFs are solved using SAT solvers, which cannot exploit the algebra of polynomials naturally.
                We propose a paradigm that bridges between ANF and CNF solving techniques:
		the techniques are applied in an iterative manner to \emph{learn facts} to augment the original problems.		
		Experiments on over 1,100 benchmarks arising from four different applications domains demonstrate that learnt facts can significantly improve runtime and enable more benchmarks to be solved.
        \end{abstract}
	
%	\begin{IEEEkeywords}
%		Algebraic Normal Form (ANF), Conjunctive Normal Form (CNF), SAT Solvers, eXtendend Linearization, ElimLin
%	\end{IEEEkeywords}
	
	\section{Introduction}
	
	Algebraic Normal Form (ANF) and Conjunctive Normal Form (CNF) are two commonly used normal forms in Boolean algebra.
	Both ANF and CNF reason about Boolean variables $x_1, \dots, x_n$ but with different Boolean operators.
	
	ANF is a \emph{system of polynomial equations} in GF(2), i.e., the  Galois field of two elements, or $\mathbb{Z}_2$.
	Each polynomial is a sum of monomials, where a monomial is a product of zero or more variables.
	%ANF is preferred by cryptologists because
	Cryptologists prefer ANF because
        it naturally encodes 
        definitions such as AES \cite{aes2001} and hash functions \cite{shs2015}. %, and %constructions such as
        %Hidden Field Equations \cite{patarin1996hidden}. % can all be naturally encoded in ANF.
	
	One approach to solving ANF is to compute the Gr\"obner basis of the system using the Buchberger's algorithm \cite{buchberger2006bruno} or its variants \cite{faugere1999new, Faugere:2002:NEA:780506.780516}.
	Efficient implementations include M4GB \cite{makarim2017M4GB}, FGb \cite{faugere2010fgb} and Magma \cite{bosma1997magma}.
	In certain systems, methods such as XL/XSL \cite{courtois2000efficient,courtois2002cryptanalysis} and ElimLin \cite{courtois2007algebraic,courtois2012elimlin} have also been shown to be effective.
	Unfortunately, ANF solvers on huge polynomial systems tend to require more memory than is feasible on most computing platforms \cite{gao2010new}.
	
	In comparison, CNF is a \emph{conjunction of clauses}.
	Each clause is a disjunction of literals, where a literal is either a Boolean variable or its negation.
	As Boolean circuits are naturally described in logical connectives, hardware verification problems are often described in CNFs \cite{prasad2005survey}.
	Some other domains using CNFs are
	software verification, industrial planning, scheduling and recreational mathematical puzzle solving. 
	
	CNFs are typically solved by SAT solvers, which use significantly less memory than the methods for ANF.
	This is primarily due to the depth-first search nature of CDCL~\cite{marques1999grasp} that most modern SAT solvers are based on.
    Many solvers build upon the small code base of MiniSat \cite{een2003extensible}, which includes the standard CDCL, variable and clause elimination~\cite{varelim}, watched literals data structures \cite{moskewicz2001chaff} and the like.
	
	ANF and CNF solving algorithms exploit different properties of the problem encoding.
	For instance, Gauss-Jordan elimination (GJE) is a natural procedure in ANF, but not in CNF;
	while conflict learning prunes the search tree in SAT solvers, but we are unaware of such learning for ANF.
	Despite %efforts at integrating GJE into SAT solvers \cite{laitinen2014extending,han2012boolean,soos2009extending} and
        the recent successes of GJE-enabled SAT solvers in counting problems \cite{meel2016constrained,Meel17}, the use of GJE-enabled solvers is not prevalent. In this context, we ask: {\em is there an alternative and easier way to combine ANF and CNF solving?}% techniques?}
	
	The primary contribution of this paper is an affirmative answer to the above question. We demonstrate a paradigm that bridges between ANF and CNF solving techniques. 
	The techniques are applied in an iterative manner to \emph{learn facts} to augment the original problems. 
	This approach is attractive when the conversion time between ANF and CNF encodings is negligible relative to the overall solving time.
	Our experiments demonstrate that our iterative approach can help us to solve more\haveSAT{ ANF}{}
	instances while spending less time.

	As a consequence of this bridge, problems can be encoded in their most natural and comprehensible manner, either in ANF or CNF, and yet draws from solving techniques in both to achieve reasonable solving performance --- this is our second contribution.
	We call our tool \textsc{Bosphorus}, the namesake of \emph{the Bosphorus bridge} connecting Europe and Asia.	
	
	In the next section, we describe the various techniques for solving ANFs and CNFs.
	Section \ref{sec:bosphorus} describes how \textsc{Bosphorus} uses these techniques. %  and also gives implementation details.
	Results on three classes of ANF problems\haveSAT{}{ and the SAT Competition 2017 benchmarks} are in section \ref{sec:experiments}.
	For notation, we use $\oplus$ for exclusive-OR (XOR) and addition in GF(2),
	$\lnot$ for negation, $\land$ for conjunction and $\lor$ for disjunction.
	We use the term \emph{polynomial} to mean \emph{polynomial equation equated to zero},
	and we will also write such equations by just stating the polynomial.

	\section{Learning Facts}
	\label{sec:techniques}
	
	Our approach iteratively extracts two types of \emph{learnt facts}: (1) linear equations $x_{i_1} +  x_{i_2} +\dots + x_{i_p} + c$ where $c$ is either zero or one; and (2) polynomials of the form $x_{i_1} x_{i_2} \dots x_{i_p} \oplus 1$.
	The former keeps the degree of the system low while the latter allows immediate deduction that $x_{i_1} = x_{i_2} = \dots = x_{i_p} = 1$.
	The rest of this section explains how \textsc{Bosphorus} obtains and uses these facts in various phases.
	
	\subsection{ANF propagation}
	\label{sec:prop}
	
	For each variable, we attempt to assign a \emph{value} (0 or 1) or an \emph{equivalent literal} by examining the polynomials involving the variable.
	A value assignment can occur in two cases.
	First, for polynomial $x$ or $x \oplus 1$, we set $x$ to the constants 0 or 1 respectively.
	Second, for polynomial $x_{i_1} x_{i_2} \dots x_{i_p} \oplus 1$, we set $x_{i_1} = x_{i_2} = \dots = x_{i_p} = 1$.
	An equivalence assignment happens if the polynomial is $x \oplus y$ or $x \oplus y \oplus 1$, in which case we set $x = y$ or $x = \lnot y$ respectively.
	These assignments are applied iteratively until a fixed point is reached.

%FIXME: add ANF propagation hack if that has been applied in the experiments.
        
	\subsection{eXtended Linearization (XL)}
	\label{sec:xl}
	
	Gauss-Jordan elimination (GJE) solves a system of linear equations by elementary row operations.
	For polynomials, one can apply GJE by treating each monomial as an independent variable --- this is known as \emph{linearization}.
	Dependence between the monomials can be re-introduced by generating more polynomial equations, a process known as eXtended Linearization (XL) \cite{courtois2000efficient}.
	We describe XL and how it is used.
	
	Given a polynomial system $S$ with $n$ variables and $m$ equations, we expand $S$ incrementally\haveSAT{, terminating when the expanded system $S'$ is too large when linearized.}{ to obtain an expanded system $S'$.}
	The expansion process selects each equation in $S$ in ascending degree order and multiplies the equation with all possible monomials up to a chosen degree $D$.
	In the case where we manage to expand $S$ fully, the expanded system will have $m \sum_{j=0}^D \binom{n}{j}$ polynomials.
	GJE is then applied on $S'$.
	
	Table \ref{tbl:xleg} shows an example of applying XL on the ANF $\lbrace x_1 x_2 \oplus x_1 \oplus 1, x_2 x_3 \oplus x_3\rbrace$, expanding up to degree ${D=1}$  monomials.
	The last three rows of Table \ref{tbl:xlgjeeg} are the facts $\lbrace x_1 \oplus 1, x_2, x_3\rbrace$ that \textsc{Bosphorus} will retain.
	
	\begin{table*}
		\newcommand\z{}
		\caption{An example of applying eXtended Linearization (XL). Zero coefficients in the matrices are suppressed; and rows corresponding to zero polynomials are omitted. The last three rows of (b) are the facts that will be retained.}
		\label{tbl:xleg}
		\begin{subtable}[t]{1.2\columnwidth}
			\caption{Expansion by degree 1 monomials}
			\begin{tabular}{ccc@{\hspace{2.5ex}}c@{\hspace{2.5ex}}c@{\hspace{2.5ex}}c@{\hspace{2.5ex}}c@{\hspace{2.5ex}}c@{\hspace{2.5ex}}c@{\hspace{2.5ex}}c@{\hspace{2.5ex}}c}\toprule
				&&\multicolumn{8}{c}{Expanded linearized system}\\\cmidrule(lr){3-10}
				Polynomial & Multiplier & $x_1 x_2 x_3$ & $x_2 x_3$ & $x_1 x_3$ & $x_1 x_2$ & $x_3$ & $x_2$ & $x_1$ & $1$\\
				\midrule
				$x_1 x_2 \oplus x_1 \oplus 1$ & $1$   & \z & \z & \z &  1 & \z & \z &  1 & 1\\
				& $x_1$ & \z & \z & \z &  1 & \z & \z & \z & \z\\
				& $x_2$ & \z & \z & \z & \z & \z &  1 & \z & \z\\
				& $x_3$ &  1 & \z &  1 & \z &  1 & \z & \z & \z\\
				$x_2 x_3 \oplus x_3$          & $1$   & \z &  1 & \z & \z &  1 & \z & \z & \z\\
				& $x_1$ &  1 & \z &  1 & \z & \z & \z & \z & \z\\
				& $x_3$ & \z &  1 & \z & \z &  1 & \z & \z & \z\\
				\bottomrule
		\end{tabular}\end{subtable}
		\hfill
		\begin{subtable}[t]{0.8\columnwidth}
			\caption{Gauss-Jordan Elimination}
			\label{tbl:xlgjeeg}
			\begin{tabular}{c@{\hspace{2.5ex}}c@{\hspace{2.5ex}}c@{\hspace{2.5ex}}c@{\hspace{2.5ex}}c@{\hspace{2.5ex}}c@{\hspace{2.5ex}}c@{\hspace{2.5ex}}c}
				\toprule
				\multicolumn{8}{c}{Linearized system after GJE}\\\cmidrule(lr){1-8}  
				$x_1 x_2 x_3$ & $x_2 x_3$ & $x_1 x_3$ & $x_1 x_2$ & $x_3$ & $x_2$ & $x_1$ & $1$\\
				\midrule
				1 & \z &  1 & \z & \z & \z & \z & \z\\
				\z &  1 & \z & \z & \z & \z & \z & \z\\
				\z & \z & \z &  1 & \z & \z & \z & \z\\
				\z & \z & \z & \z &  \textbf{1} & \z & \z & \z\\
				\z & \z & \z & \z & \z &  \textbf{1} & \z & \z\\
				\z & \z & \z & \z & \z & \z &  \textbf{1} &  \textbf{1}\\\\\bottomrule
		\end{tabular}\end{subtable} 
	\end{table*}
	
        \haveSAT{}{
          Applying XL on the entire ANF often requires considerable memory and time.
          To avoid this, we uniformly subsample the polynomials from the ANF to obtain an $m'$-by-$n'$ linearized system $S$ such that ${m'n'\gtrsim 2^M}$, for a fixed parameter $M$.
          Moreover, $S$ is incrementally expanded only until the system size is approximately $2^{M+\delta M}$, for a parameter $\delta M$.

          We employ XL in this manner because our primary purpose is not to solve the system but to learn facts to augment it.
          We also employ ElimLin and SAT solver in the same spirit.
        }
        
	\subsection{ElimLin}
	\label{sec:el}
	
	ElimLin \cite{courtois2007algebraic} is an algorithm that iterates through the following three steps until fixed point:
	(1) apply GJE on the linearization of the polynomial system $S$;
	(2) gather linear equations and remove them from $S$, yielding $S'$; and
	(3) for each linear equation $\ell$, pick, say, a variable from $\ell$ that occurs in the least number equations in $S'$, and eliminate that variable from $S'$ using $\ell$.
	The resultant system $S''$ is free of linear equations.
	The process is repeated from step (1) using $S''$ as $S$ until there are no more linear equations after applying GJE.
	
	Consider the ANF $\lbrace x_1 \oplus x_2 \oplus x_3, x_1 x_2 \oplus x_2 x_3 \oplus 1\rbrace$.
	As step (1) does not affect the system, $x_1 \oplus x_2 \oplus x_3$ remains the only linear equation in step (2).
	If we choose to substitute $x_1$ by $x_2 \oplus x_3$ in step (3), the ANF becomes the single equation $(x_2 \oplus x_3) x_2 \oplus x_2 x_3 \oplus 1$.
	By right-distributing the first conjunction over the first XOR and then replacing the XOR of $x_2 x_3$ with itself by zero, 
	this equation simplifies to $x_2 \oplus 1$.
	Assigning $x_2 = 1$ and performing ANF propagation on the original ANF, $x_1 x_2 \oplus x_2 x_3 \oplus 1$ becomes $x_1 \oplus x_3 \oplus 1$, and the ANF propagation can deduce the equivalence $x_1 = \lnot x_3$.
	
        \haveSAT{}{
          Similar to XL, we apply ElimLin on a random subset of polynomials that has linearized size of approximately $2^M$.
        }

        \subsection{Conflict-bounded SAT solving}
	\label{sec:sat}
	
	With a CNF equivalent of the ANF,
	%\textsc{Bosphorus} calls
        we call
        a SAT solver that has conflict-driven clause learning \cite{marques1999grasp}.
	The solver is allowed up to a pre-determined number $C$ of conflicts to solve the system.
	We bound the solver using
        use a conflict budget instead of a time budget for replicability of experiments.
        
	Due to this budget, the solver will surely terminate with one of these three cases:
	(1) unsatisfiable;
	(2) satisfiable, giving an assignment; or
	(3) undecidable within the limit.
	In case (1), \textsc{Bosphorus} appends
        %we append
        the contradictory equation ${1 = 0}$ to the system --- this is the learnt fact by the SAT solver.
	In cases (2) and (3),
        \textsc{Bosphorus} extracts
        %we extract
        linear equations from learnt clauses --- of particular interest are linear equations from the unit and binary clauses  because they immediately yield value and equivalence assignments.

	\subsection{Example}
	
	Consider the ANF
	\begin{gather}
	\begin{aligned}
	&x_1 x_2 \oplus x_3 \oplus x_4 \oplus 1,&
	&x_1 x_2 x_3 \oplus x_1 \oplus x_3 \oplus 1,\\
	&x_1 x_3 \oplus x_3 x_4 x_5 \oplus x_3,&
	&x_2 x_3 \oplus x_3 x_5 \oplus 1,\\
	&x_2 x_3 \oplus x_5 \oplus 1.
	\end{aligned}
	\label{eqn:egproblem}
	\end{gather}
	XL with $D = 1$ on this system learns the facts
	$x_2 x_3 x_4 \oplus 1$,
	$x_1 x_3 x_4 \oplus 1$,
	$x_1 \oplus x_5 \oplus 1$,
	$x_1 \oplus x_4$, 
	$x_3 \oplus 1$, and
	$x_1 \oplus x_2$.
\cut{	
	\begin{align*}
	&x_2 x_3 x_4 \oplus 1, &
	&x_1 x_3 x_4 \oplus 1, &
	&x_1 \oplus x_5 \oplus 1,\\
	&x_1 \oplus x_4, &
	&x_3 \oplus 1,&
	&x_1 \oplus x_2.
	\end{align*} }
	For ElimLin, its initial GJE --- step (1) in section~\ref{sec:el} --- gives four distinct linear equations: $x_1 \oplus x_5 \oplus 1$; $x_1 \oplus x_4$; $x_3 \oplus 1$; and $x_1 \oplus x_2$. 
	After substituting $x_5$ by $x_1 \oplus 1$, $x_4$ by $x_1$, $x_3$ by 1 and $x_2$ by $x_1$, ElimLin learns $x_1 \oplus 1$.
	Converting to CNF using Karnaugh map (section~\ref{sec:anf2cnf}) creates one auxiliary variable for $x_1 x_2$.
	Boolean constraint propagation in the SAT solver then gives
%	Using Karnaugh map conversion (section~\ref{sec:anf2cnf}), only one auxiliary variable $x_1 x_2$ is created in the CNF given to the SAT solver.
%	Boolean constraint propagation then gives
	$x_2 \oplus 1$,
	$x_4 \oplus 1$,
	$x_5$, and
	$x_1 x_2 \oplus 1$.
\cut{	\begin{align*}
	&x_2 \oplus 1,&
	&x_4 \oplus 1,&
	&x_5,&
	&x_1 x_2 \oplus 1.
	\end{align*} }
	
	ANF propagation using the above facts obtained from XL, ElimLin and SAT solver simplifies the system into
	\begin{align}
	&x_1 \oplus 1,&
	&x_2 \oplus 1,&
	&x_3 \oplus 1,&
	&x_4 \oplus 1,&
	&x_5.
	\label{eqn:eganf}
	\end{align}
	This effectively solves the system to its unique satisfying assignment $x_1 = x_2 = x_3 = x_4 = 1$ and $x_5 = 0$.
	
	Observe that ANF propagation after the XL step would have led to \eqref{eqn:eganf} without the need for either ElimLin or SAT solver.
	Nevertheless, the above example illustrates that each can derive different learnt facts:
	XL gives the value assignment for $x_3$, ElimLin gives that for $x_1$, and the SAT solver learns the remaining assignments.
	To make full use of these different learnt facts, \textsc{Bosphorus} is designed to perform ANF propagation when learnt facts are produced after every step.

	\section{\textsc{Bosphorus}}
	\label{sec:bosphorus}
	This section details the workflow and the data structures of \textsc{Bosphorus}, and the approaches to convert between ANFs and CNFs.
	%We also list the software packages that are used. % in implementing \textsc{Bosphorus}.
        The source code is available at \url{https://github.com/meelgroup/bosphorus}.
	
	\subsection{Workflow}
	
\begin{figure}
		\centering
		\footnotesize
		\begin{tikzpicture}[node distance = 2cm]
		\node [document] (toanf) {Convert\\to ANF};
		\node [input, left=0.5cm of toanf] (inanf) {Problem\\description};
		\node [decision, aspect=3, below=0.5cm of toanf] (loop) {Fixed Point};
		\node [block, below=0.4cm of loop] (xl) {XL};
		\node [block, left=1cm of xl] (el) {ElimLin};
		\node [block] (sat) at (el |- loop) {SAT Solver};
		\node [document, right=1cm of loop] (outanf) {Processed\\ANF and CNF};
		
		\draw [->] (inanf) -- (toanf);
		\draw [->, dashed] (toanf) -- (loop);
		\draw [->] (loop) --node[anchor=east] {no} (xl);
		\draw [->, dashed] (xl) -- (el);
		\draw [->, dashed] (el) -- (sat);
		\draw [->, dashed] (sat) -- (loop);
		\draw [->] (loop) --node[anchor=south] {yes} (outanf);
		\end{tikzpicture}
		\caption{\textsc{Bosphorus}'s flow. A dashed arrow means ANF propagation is applied.}
		\label{fig:Bosphorus}
	\end{figure}
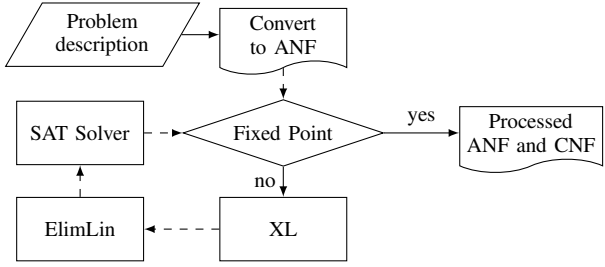

	\textsc{Bosphorus} takes a problem encoded in ANF and produces a processed ANF and CNF after performing an XL--ElimLin--SAT-solver \emph{fact-learning loop} until the fixed point when no further learnt facts are produced.
	ANF propagation is performed on the input ANF and whenever learnt facts are produced.
	Fig.~\ref{fig:Bosphorus} shows the overall workflow.
	
	Internally within \textsc{Bosphorus}, the problem is represented as an ANF polynomial system,
	and only ANF propagation modifies and replaces this master copy.
	Each of the other techniques --- XL, ElimLin and SAT solver --- operates on a copy of the ANF,
	and learnt facts are extracted and then added onto the master copy if not already there.

	If the equation ${1 = 0}$ is detected, \textsc{Bosphorus} terminates and returns UNSAT.
	%If a satisfying solution is found by the SAT solver, \textsc{Bosphorus} stores the solution. 
	If the SAT solver finds a satisfying solution, \textsc{Bosphorus} stores the solution. 
	This solution is not used to simplify the ANF because it may not be unique. %there may be more than one solution.

	\subsection{Data structures}
	
	\textsc{Bosphorus} stores the system of equations in the ANF description as a list of Boolean polynomials.
	For each variable, we track
	%\begin{itemize}
	(i) its value, as either 0, 1, or undetermined;
	(ii) its equivalence literal; and
	(iii) its occurrence list.
	%\end{itemize}
	
	The default equivalence literal for each variable is the variable itself and may change as \textsc{Bosphorus} proceeds.
	For example, the equivalence literal of $x_i$ may be switched to $\lnot x_j$ to encode $x_i = \lnot x_j$.
	
	Occurrence list is an optimization technique from the SAT literature \cite{moskewicz2001chaff,zhang1997sato}.
	Here, \textsc{Bosphorus} tracks the list of polynomials that each variable occurs in.
	For example, updates to $x_1$ in \eqref{eqn:egproblem} do not involve processing the last two equations. % that do not involve $x_1$.
	The time saved can be significant for large polynomial systems.
	
	\subsection{ANF to CNF conversion}
	\label{sec:anf2cnf}
	
	CNF is used by the SAT solver within \textsc{Bosphorus}, and it is also an output.
	To convert ANF to CNF, we introduce an auxiliary CNF variable on-the-fly for each ANF monomial, and we maintain a bi-directional map for such variables.

	\textsc{Bosphorus} handles determined variables, equivalences, and polynomials differently in the conversion.
	Determined variables are added as unit clauses, while an equivalence such as $x_i = \lnot x_j$ is represented in CNF by $(x_i \lor x_j) \land (\lnot x_i \lor \lnot x_j)$.
	For a polynomial, it is first re-expressed as shorter ones by introducing auxiliary variables.
	The number of terms in the shorter polynomials is parameterized by an XOR-cutting length $L$,  
	Then, each of these shorter polynomials is converted to CNF using either of the following two approaches:
	\begin{enumerate}
		\item If the polynomial is $K$-variate, we use the Karnaugh map to yield the minimal clause representation while reducing the number of auxiliary variables used.
		Because computing the Karnaugh map scales exponentially with the number of variables, the Karnaugh parameter $K$ is kept low to ensure reasonable conversion time.
		\item If the polynomial involves more than $K$ variables, we apply a transformation \emph{\`{a} la} Tseitin encoding \cite{tseitin1983complexity}.
		Each polynomial of length $l \leq L$ is treated as an XOR clause of independent terms and converted to CNF clauses by enumerating through all possible $2^l$ terms.
	\end{enumerate}
	
	Although the Karnaugh map approach is less flexible, it can yield a more compact conversion than the Tseitin-based approach.
	Consider the polynomial equation ${x_{1} x_{3} \oplus x_{1} \oplus x_{2} \oplus x_{4} \oplus 1 = 0}$.
	Fig.~\ref{fig:anftocnf} shows possible CNF representations via both approaches.
	Using the Karnaugh map shown in Fig.~\ref{fig:kmap}, one can derive a more compact CNF system that directly deals with the variables involved.
	In comparison, the Tseitin-based approach creates a new CNF variable $x_5$ and encode ${x_5=x_{1} x_{3}}$ using three CNF clauses.
	
	\begin{figure}
		\footnotesize
		\newcommand\h{\hphantom{\lnot}}
		\centering
		\begin{minipage}{0.45\columnwidth}
			\begin{flalign*}
			\h x_1 \lor \h x_2 \lor \h x_4\\
			\lnot x_1 \lor \lnot x_2 \lor \h x_3 \lor \h x_4\\
			\h x_2 \lor \lnot x_3 \lor \h x_4\\
			\lnot x_1 \lor \h x_2 \lor \h x_3 \lor \lnot x_4\\
			\h x_1 \lor \lnot x_2 \lor \lnot x_4\\
			\lnot x_2 \lor \lnot x_3 \lor \lnot x_4\\
			\end{flalign*}
		\end{minipage}
		\hfill
		\begin{minipage}{0.45\columnwidth}
			\begin{flalign*}
			\h x_1 \lor \lnot x_5\\
			\h x_3 \lor  \lnot x_5 \\
			\lnot x_1 \lor \lnot x_3 \lor \h x_5 \\
			\h x_1 \lor \h x_2 \lor \h x_4 \lor \h x_5\\
			\lnot x_1 \lor \lnot x_2 \lor \h x_4 \lor \h x_5\\
			\lnot x_1 \lor \h x_2 \lor \lnot x_4 \lor \h x_5\\
			\h x_1 \lor \lnot x_2 \lor \lnot x_4 \lor \h x_5\\
			\lnot x_1 \lor \h x_2 \lor \h x_4 \lor \lnot x_5\\
			\h x_1 \lor \lnot x_2 \lor \h x_4 \lor \lnot x_5\\
			\h x_1 \lor \h  x_2 \lor \lnot x_4 \lor \lnot x_5\\
			\lnot x_1 \lor \lnot x_2 \lor \lnot x_4 \lor \lnot x_5\\
			\end{flalign*}
		\end{minipage}
		\caption{ANF-to-CNF conversions of polynomial $x_{1} x_{3} \oplus x_{1} \oplus x_{2} \oplus x_{4} \oplus 1$. (Left) Karnaugh map conversion (6 CNF clauses); (Right) Tseitin-based conversion (11 CNF clauses).}
		\label{fig:anftocnf}
	\end{figure}
	
	\begin{figure}
		\footnotesize
		\centering
		\begin{Karnaugh}
			\contingut{1,0,1,0,0,1,0,1,0,1,1,0,1,0,0,1}
			\implicantcostats{0}{2} % -A /\ -B /\ -D
			\implicantdaltbaix{2}{10} % -B /\ C /\ -D
			\implicant{5}{7} % -A /\ B /\ D
			\implicant{7}{15} % B /\ C /\ D
			\implicant{12}{12} % A /\ B /\ -C /\ -D
			\implicant{9}{9} % A /\ -B /\ -C /\ D
		\end{Karnaugh}
		\caption{Karnaugh map of polynomial $x_{1} x_{3} \oplus x_{1} \oplus x_{2} \oplus x_{4} \oplus 1$.}
		\label{fig:kmap}
	\end{figure}
	
        At present, any auxiliary variable introduced in the conversion process does not participate in the learnt facts.
        
	\subsection{CNF to ANF conversion}
	
	%Many verification problems are naturally described in CNF \cite{prasad2005survey}.
    \textsc{Bosphorus} can be used as a CNF preprocessor, though its main use-case is that of solving problems represented in ANF. When used as a CNF preprocessor, \textsc{Bosphorus} obtains an equivalent ANF in the following manner \cite{hsiang1985refutational}:
	\begin{enumerate}
		\item Each CNF variable is assigned a unique ANF variable;
		\item Each clause is converted to a polynomial via product of negated literals.
	\end{enumerate}
	For instance, the polynomial for the clause $\lnot x_1 \lor x_2$ is  $(x_1)(x_2 \oplus 1) = x_1 x_2 \oplus x_1$.
	The resultant polynomial degree is the number of literals in each clause.
	More importantly, if a clause has $n$ positive literals, there will be $2^n$ terms in the polynomial.
	To prevent such cases, %we apply a technique similar to the ANF-to-CNF conversion in section~\ref{sec:anf2cnf}:
	we re-express the clause as a set shorter of clauses by introducing auxiliary variables \emph{\`{a} la} converting a $k$-SAT to 3-SAT.
	We limit the number of positive literals within each of the shorter clauses to $L'$, called the clause-cutting length.
	Each of the shorter clauses is then converted to polynomials as outlined above.
    
    %This CNF-to-ANF conversion is a trivial one, unlike \cite{grobner-preprocess} where the conversion to ANF is highly sophisticated. Instead, we perform a trivial conversion and then use sophisticated techniques to simplify the problem on the ANF level.
    This CNF-to-ANF conversion is trivial, unlike that in \cite{grobner-preprocess}; sophisticated techniques are then applied to simplify the problem on the ANF level.
	In this use-case,
	converting problem from CNF to ANF and back to CNF give  a suboptimal description of the original problem.
	Hence, %when given a problem in CNF,
        \textsc{Bosphorus} returns the original CNF in addition to the one converted from its internal ANF representation, which includes the learnt facts.

	\subsection{Implementation}
	
	\textsc{Bosphorus} uses the following existing work:
	\begin{LaTeXdescription}
		\item[\normalfont\emph{PolyBoRi\cite{brickenstein2009polybori}}] To store and manipulate Boolean polynomials.
		\item[\normalfont\emph{M4RI \cite{albrecht2012M4RI,albrecht2011M4RI}}] For efficient Gauss-Jordan elimination on Boolean matrices, necessary for XL and ElimLin.
		\item[\normalfont\emph{CryptoMiniSat5 \cite{soos2016cryptominisat}}] This is a SAT solver equipped with conflict-driven clause learning.
		To extract learnt facts from this solver, we modify version 5.6.3 of the solver to exposed its APIs that extract \emph{linear equations}.
		%Alternatively, one may also consolidate all the learnt clauses and post-process them without the SAT solver.
		\item[\normalfont\emph{ESPRESSO \cite{brayton1984logic}}] For Karnaugh map simplification \cite{karnaugh1953map}.
		While ESPRESSO is a heuristic logic minimizer, it is fast and often yields close-to-optimum representations.
	\end{LaTeXdescription}

	\section{Experiments and Results}
	\label{sec:experiments}
	
	\haveSAT{%
		We run experiments on three types of the problem described in ANFs: round-reduced AES cipher, round-reduced SIMON cipher, and weakened Bitcoin nonce finding.
	}{%
		We run experiments on three classes of problem described in ANFs and a set of problems in CNFs.
		The ANF problems are round-reduced AES cipher, round-reduced SIMON cipher and weakened Bitcoin nonce finding,
		while the CNF problems consist of a wide variety from the SAT Competition 2017 \cite{balyo2017sat}.
	}%
	These problems are detailed in the appendix.
	The experiments are conducted on a single Intel Xeon E5-2670v2 2.50GHz processor core. 
	Each ANF\haveSAT{}{ or CNF} is passed to \textsc{Bosphorus}, which, after learning facts using the XL--ElimLin--SAT-solver loop together with ANF propagation, will give a CNF that includes the learnt facts.
	A SAT solver is then used to solve the processed CNF eventually. Note that the most efficient off-the-shelf ANF solver, M4GB, has such  a high memory footprint that it times out on all the instances. 
%	This is done\haveSAT{}{ also for the ANF problems} because M4GB \cite{makarim2017M4GB}, which is the most memory efficient ANF solver that we know, uses more memory than is available on our platforms for these problems.
	
	%As a comparison,
        We also pass the instances to the SAT solvers directly without learning facts but only converting to CNFs using \textsc{Bosphorus}\haveSAT{}{ if needed}.
	We also evaluate with three different SAT solvers for the eventual solving: a minimalistic SAT solver MiniSat \cite{een2003extensible}, a high-performance SAT solver Lingeling \cite{biere2017cadical}, and CryptoMiniSat5 \cite{soos2016cryptominisat}, which natively performs Gauss-Jordan elimination.\footnote{The versions used are 2.2, bcj-78ebb86-180517 and 5.6.3 respectively.}
	We report the PAR-2 score \cite{balyo2017sat} and the number of solved instances.
	The PAR-2 score is the sum of runtimes for solved instances and twice the timeout for unsolved instances, and a lower score is better.
	
	\haveSAT{
          For the \textsc{Bosphorus}'s workflow, we use the following parameters: XL expansion degree ${D = 1}$, Karnaugh parameter ${K = 8}$, cutting lengths ${L = L' = 5}$, and SAT-solver conflict budget ${C = 100,000}$.
        }{
	  For the \textsc{Bosphorus}'s workflow, we use the following parameters: XL and ElimLin subsampling parameter ${M=30}$, XL expansion allowance ${\delta M=4}$ and degree ${D = 1}$, Karnaugh parameter ${K = 8}$, cutting lengths ${L = L' = 5}$, and SAT-solver conflict budget starting from ${C = 10,000}$, increasing up to $100,000$ in increments of $10,000$ when the learnt clauses from the SAT-solver produce no new learnt facts.}
	\haveSAT{Since all the problems are satisfiable with (most probably) a single solution, we have also made \textsc{Bosphorus} exit the loop if the SAT solver finds a satisfying assignment.}{
		Moreover, we make \textsc{Bosphorus} exit the loop and provide the solution if the SAT solver finds a satisfying assignment.}
	We limit the total time used for each instance to 5,000 seconds\haveSAT{.}{, with \textsc{Bosphorus} given at most 1,000 seconds.}

	\begin{table}
		\centering
		\caption{%All instances in benchmarks are satisfiable ANFs. To convert to CNF, \textsc{Bosphorus} is used, with and without the fact-finding loop. A total timeout of 5000 seconds is enforced.
		    The PAR-2 score is shown in thousands (lower is better) with, in parenthesis, the number of solved satisfiable instances plus (if any) the number of solved unsatisfiable instances.
		  For each problem set, there are two rows of results: the first without using {\Bosphorus} (labeled \emph{w/o} in the second column) and the second with (labeled \emph{w}).
                  The better of the two is in bold, with preference to the number of solved instances.
%			There are 500 instances for the AES SR(1,4,4,8) problem set\haveSAT{and 50 instances for the other problem sets.}{, 50 instances for the Simon and Bitcoin problem sets, and 25 instances for the SAT 2017 problem set.}
		}
		\label{tab:results}
                % Note on results
                % Simon-10: ANF propagation without further learning hurts CMS5; Just conversion is better for CMS5!!               
		\begin{tabular}{llrrr}
			\toprule
			Problem & & \multicolumn{1}{c}{MiniSat} & \multicolumn{1}{c}{Lingeling} & \multicolumn{1}{c}{CryptoMiniSat5} \\
			\midrule
			SR-[1,4,4,8]& w/o &         4372 (\hphantom{4}89)  &          532 (500)  & \textbf{504 (500)}\\
			(500)       & w   & \textbf{1099 (489)} & \textbf{518 (500)} &         507 (500) \\[1.5ex]
			Simon-[8,6] & w/o &   \textbf{1 (50)} &  \textbf{0 (50)}  &  \textbf{0 (50)}\\
			(50)        & w   &           3 (50)  &          3 (50)   &          3 (50) \\[1ex]
			Simon-[9,7] & w/o &         324 (22)  &  \textbf{0 (50)}  &  \textbf{2 (50)}\\
			(50)        & w   &  \textbf{15 (50)} &         14 (50)   &         14 (50) \\[1ex]
			Simon-[10,8]& w/o &         500 (\hphantom{3}0)  &         31 (50)   &         45 (50) \\
			(50)        & w   & \textbf{231 (34)} & \textbf{29 (50)}  & \textbf{44 (50)} \\[1.5ex]
			Bitcoin-[10]& w/o &   \textbf{4 (50)} &   \textbf{9 (50)} &    \textbf{8 (50)}\\
			(50)        & w   &          23 (50)  &          23 (50)  &           24 (50)\\[1ex]
			Bitcoin-[15]& w/o & \textbf{146 (43)} & \textbf{185 (39)} &          169 (40) \\
			(50)        & w   &         171 (42)  &         220 (34)  &  \textbf{176 (41)}\\[1ex]
			Bitcoin-[20]& w/o &         493 (~1)  &         475 (~3)  &          486 (~2) \\
			(50)        & w   & \textbf{482 (~2)} & \textbf{471 (~4)} &  \textbf{477 (~3)} \\[1.5ex]
			SAT-2017    & w/o &         2105 (75+38)  &         2006 (70+56)  &         1764 (89+63)\\ 
			(310)       & w   & \textbf{2153 (72+42)} & \textbf{2070 (70+57)} & \textbf{1674 (98+77)}\\[1ex]
			SAT-2017    & w/o & 2045 (15+\phantom{1}7)  &         1738 (29+26)  &         1689 (30+32)\\
			(219)       & w   & \textbf{1981 (18+11)} & \textbf{1756 (29+27)} & \textbf{1543 (40+46)}\\
			\bottomrule
		\end{tabular}
	\end{table}
	We only present results for selected benchmarks in Table~\ref{tab:results}. The first column represents the class of benchmarks followed by the number of instances in parenthesis. For each problem class, we have two rows of results: the first without using {\Bosphorus} and the second with. The third, fourth and fifth columns specify the PAR-2 score (in thousands) for MiniSat, Lingeling, and CryptoMiniSat5 respectively.
        %PAR-2 scheme, that is, penalized average runtime, used in SAT-2017 Competition~\cite{SAT2017Proceedings}, assigns a runtime of two times the time limit (instead of a not solved status) for each benchmark not solved by a solver.
        The PAR-2 score for the case of using {\Bosphorus} includes time taken by {\Bosphorus}.
        The Simon, Bitcoin and SAT-2017 benchmark classes are listed in increasing difficulty.
	
	 For the instances from SR-[1,4,4,8], {\Bosphorus} allows a significantly more solved instances for MiniSat, and it provides similar PAR-2 scores for Lingeling and CryptoMiniSat5 even while including its overhead.
         Similar observations can be made for the harder Simon instances, though the overhead of {\Bosphorus} is now clearly visible in Simon-[9,7].
         With Bitcoin, {\Bosphorus} does not always help, but the effect of its overhead to the PAR-2 scores diminishes with the harder instances.
         One way to study when {\Bosphorus} helps is to run it with different parameters.
         For the SAT-2017 CNF instances, {\Bosphorus} does provide useful information to the solvers, especially for the UNSAT instances.

         \section{Discussion}
         
%         %Overall, we see that usage of {\Bosphorus} enables us to solve more benchmarks in lower runtime.
%         To put our contribution in context we quote two leading SAT developers Professors Audemard and Simon: % to put our results in perspective:
%``We must also say, as a preliminary, that improving SAT solvers is often
%a  cruel  world.
%%To  give  an  idea,
%...
%improving  a  solver  by  solving  at  least
%ten more instances ... %(on a fixed set of benchmarks of a competition) is
%is generally showing a critical new feature.'' % In general, the winner of a competition is decided based on a couple of additional solved benchmarks.''

While {\Bosphorus} can be used as a CNF preprocessor, %as is the case for some instances of the SAT-2017 problems,
it is in fact a flexible reasoning framework on Boolean or GF(2) variables in the following sense.
First, for satisfiable problems, the SAT solver collapse onto \emph{one} solution, while {\Bosphorus} can continuously constrain the solution space without committing to one particular solution.
Second, for unsatisfiable problems, the conclusion can be reached by either any of the ANF techniques giving $1=0$ or by the SAT solver giving UNSAT.
Third, any of the solving techniques in the workflow can be improved with minimal impact on the other techniques because the retained facts do not increase the complexity of the equations.
Fourth, it is relatively easy to include new solving techniques by plugging them as components into the workflow, for example, lookahead SAT solvers \cite{SAThandbook} and Buchberger's algorithm \cite{buchberger2006bruno}.
In fact, using the Buchberger's algorithm as a preprocessor for SAT solving has previously been proposed \cite{grobner-preprocess}, but, with {\Bosphorus}, it may now be applied in an iterative manner together with other solving techniques.
	
To conclude, we have proposed and implemented a tool named \textsc{Bosphorus} that iteratively applies eXtended linearization, ElimLin and conflict-bounded SAT solving together with ANF propagation in order to learn additional facts to augment the original problem.
The experiments on selected ANF\haveSAT{}{ and SAT} problems have shown that this approach can help solve more problems in a shorter time, particularly for the harder instances.

\cut{
       \section{Conclusion}
	
	We have proposed and implemented a tool named \textsc{Bosphorus} that iteratively applies eXtended linearization, ElimLin and conflict-bounded SAT solving together with ANF propagation in order to learn additional facts to augment the original problem.
	The experiments on selected ANF\haveSAT{}{ and SAT} problems have shown that this approach can help solve more problems in a shorter time, particularly for the harder instances.

	%%% Remove future work if lack of space
	This paper is a proof of concept, and there are multiple directions to extend and improve it.
	On a more global level, we may investigate other kinds of facts can be usefully propagated throughout the system; and we may include other solving techniques.
	On a more local level, each of the techniques has choices that can be further studied:
	%for eXtended linearization, instead of expanding the original system by all monomials, one may choose a subset of such monomials \cite{courtois2000efficient};
	for example in ElimLin, the choices are the variables to eliminate in the linear equations.
	%and the CNF-to-ANF conversion can be improved to facilitate fact-finding.
	
}	
	\anon{}{
	\noindent\paragraph*{Acknowledgments}
		{\small
		We thank Joshua Wong for running the M4GB experiments, Sze Ling Yeo for her Simon encoding, and Volodymyr Skladanivskyy for his SHA256 encoding. We also thank Karsten Nohl at Security Research Labs for supporting some of this work. This work was supported in part by NUS ODPRT Grant R-252-000-685-133 and AI Singapore Grant R-252-000-A16-490.%\footnote{\url{https://bitbucket.org/vsklad/cgen/src}.}.
	}
}
	
	\bibliographystyle{IEEEtran}
	\bibliography{bosphorus}
	
	\appendix
	
	\subsection{Round-reduced AES cipher --- 500 instances}
	
	We obtain a parameterized ANF encoding of AES \cite{cid2005small} from SageMath \cite{sagemath}.
	Using parameters $(n,r,c,e) = (1,4,4,8)$, we generate 500 ANF instances for 1-round AES. First, 500 random pairs of plaintext $(P)$ and key $(K)$ bits are generated and simulated to yield the corresponding ciphertext $(C)$ bits.
	%Then, the generated $(P,C)$ bits are added as $x$ or $x \oplus 1$ into the base ANF polynomial system while the $K$ bits are added as comments for verification of solutions.
	The resultant ANF has 800 variables and 1120 equations --- 864 equations and 256 bit assignments from $(P,C)$.
	%Due to encoding of \cite{cid2005small}, we skip a $(P,K)$ pair whenever zero inversions occur.

	\subsection{Round-reduced Simon cipher --- 50 instances per $(n,r)$}
	
\begin{figure}
		\centering
		\footnotesize
		\begin{tikzpicture}
		\node[draw, rectangle, minimum width=1cm] (xi1) {$x_{i+1}$};
		\node[draw, rectangle, minimum width=1cm, right=4 cm of xi1] (xi) {$x_{i}$};
		\node[draw, rectangle, minimum width=1cm, below=3 cm of xi1] (xi2) {$x_{i+2}$};
		\node[draw, rectangle, minimum width=1cm, right=4 cm of xi2] (xi1x) {$x_{i+1}$};
		\node[draw, circle, inner sep=2pt, right=of xi1.south east, anchor=north west] (S1) {$S^1$};
		\node[draw, circle, inner sep=2pt, below=0.1cm of S1] (S8) {$S^8$};
		\node[draw, circle, inner sep=2pt, below=0.1cm of S8] (S2) {$S^2$};
		\node[draw, circle, inner sep=2pt, right=of S1.south east, anchor=north west] (amp) {$\&$};
		\node at (amp -| xi) (xor1) {\Large$\oplus$};
		\node at (S2 -| xor1) (xor2) {\Large$\oplus$};
		\node [below=0.2cm of xor2] (xor3) {\Large$\oplus$};
		
		\coordinate[above=0.18cm of xor1.center] (xor1p);
		\coordinate[above=0.18cm of xor2.center] (xor2p);
		\coordinate[above=0.18cm of xor3.center] (xor3p);
		\coordinate[left=0.18cm of xor1.center] (xor1pp);
		\coordinate[left=0.18cm of xor2.center] (xor2pp);
		\coordinate[right=0.18cm of xor3.center] (xor3pp);
		
		\coordinate[above=0.2cm of xi1x] (xi1xp);
		\coordinate[above=0.2cm of xi2] (xi2p);
		\coordinate[above=0.6cm of xi2] (xi2pp);
		\coordinate[above=0.6cm of xi1x] (xi1xpp);
		
		\draw[->] (xi1) |- (S1);
		\draw[->] (xi1) |- (S8);
		\draw[->] (xi1) |- (S2);
		\draw[->] (xi1) -- (xi2pp) -- (xi1xp) -- (xi1x);
		\draw[->] (xi) -- (xor1p);
		\draw[->] (xor1.center) -- (xor2p);
		\draw[->] (xor2.center) -- (xor3p);
		
		\draw[->] (S1) -- (amp);
		\draw[->] (S8) -- (amp);
		\draw[->] (amp) -- (xor1pp);
		\draw[->] (S2) -- (xor2pp);
		
		\draw[->] (xor3.center) -- (xi1xpp) -- (xi2p) -- (xi2);
		\draw[<-] (xor3pp) -- ++(right:20pt) node[anchor=west] {$k_i$};
		\end{tikzpicture}
		\caption{One Fiestel round of Simon cipher. Diagram from \cite{beaulieu2015simon}.}
		\label{fig:simonround}
	\end{figure}
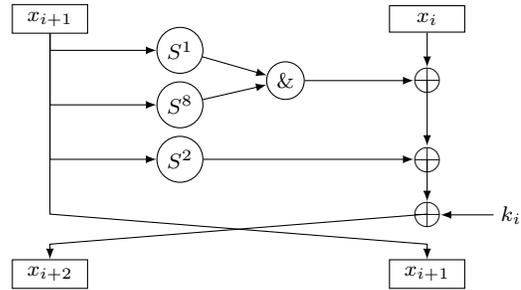

	Simon \cite{beaulieu2015simon} is a family of lightweight Feistel-based block ciphers.
	The round functions are described in conjunction and exclusive-OR of bits, allowing a straightforward ANF encoding; see Fig.~\ref{fig:simonround}. This set of benchmarks are reduced rounds Simon32/64 with multiple plaintext-ciphertext pairs encoded under the same randomly generated secret key.
	%See Appendix \ref{apdx:simon} for details on encoding Simon32/64 in ANF.
	
	Simon32/64 takes a 32-bit plaintext $(P)$ and a 64-bit key to return a 32-bit ciphertext. For each instance, we generate $n \leq 17$ plaintexts with low hamming distance as per the Similar Plaintexts/Random Ciphertexts (SP/RC) setting in \cite{courtois2014combined}.
	Concretely, the first plaintext $P_1$ is uniformly sampled while we toggle the $i\text{th}$ in the right-half of $P_1$, for $i \in \{2, \dots, n\}$.
	%flip the $(16+i-1)$-th bit of $P_1$ to obtain $P_i$, for $i \in \{2, \dots, n\}$.
	%We choose to toggle a bit in the right-half of $P_1$ because the Fiestel nature of Simon ensures that the left-half of the plaintext remains unchanged in first round. 
	This set of problems is parameterized by $(n, r)$, where $n$ is the number of plaintexts, and $r$ is the number of rounds.

%%% CUT

	\subsection{Cryptographic hash functions --- 50 instances per $k$}
	
%Cryptographically secure hash functions are traditionally used to ensure message integrity and verification of passwords.
	Recently, Cryptographically secure hash functions have been used to serve as proof-of-work in blockchains and cryptocurrencies, of which Bitcoin is an example. %such as Bitcoin and Ethereum.
	Bitcoin \cite{nakamoto2008bitcoin} uses SHA256, a hash function in the SHA-2 hash family \cite{shs2015}.

\begin{figure}
		\centering
		\footnotesize
		\begin{tikzpicture}[x=25, y=12]		
		\draw (0,0) rectangle (6,1) node[pos=0.5] {Message $M$};
		\draw (6,0) rectangle (7.5,1) node[pos=0.5] {Nonce};
		\draw (7.5,0) rectangle (8,1) node[pos=0.5] {$1$};
		\draw (8,0) rectangle (10,1) node[pos=0.5] {$|M|$};
		
		\draw [decorate,decoration={brace, amplitude=8pt}] (0,1) -- (8,1) node [] {};
		\draw [decorate,decoration={brace, amplitude=8pt}] (8,1) -- (10,1) node [] {};
		\draw (4,1.8) node [align=center] {448 bits};
		\draw (9,1.8) node [align=center] {64 bits};
		
		\draw [decorate,decoration={brace, amplitude=8pt, mirror}] (0,0) -- (6,0) node [black, midway, anchor=north, yshift=-8pt] {Randomly fixed 415 bits};
		\draw [decorate,decoration={brace, amplitude=8pt, mirror}] (6,0) -- (7.5,0) node [black, midway, anchor=north, yshift=-8pt, text width=1cm, align=center] {32-bit nonce};
		\draw [decorate,decoration={brace, amplitude=8pt, mirror}] (8,0) -- (10,0) node [black, midway, anchor=north, yshift=-8pt, text width=2cm, align=center] {Size of $M$ in binary};
		\end{tikzpicture}
		\caption{Our nonce-finding setup.}
		\label{fig:bitcoin}
	\end{figure}
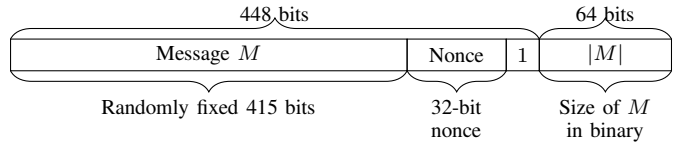

	We consider a \emph{weakened} version of the Bitcoin block hashing algorithm. %\footnote{\url{https://en.bitcoin.it/wiki/Block_hashing_algorithm}.}.
        Let $M$ be a 512-bit input message, and $H$ be a 256-bit hash output.
	We randomly set the first 415 bits of $M$, allow the next 32-bit nonce to be free (but to be determined), and pad according to SHA padding (add `1', then encode $|M| = 448$ in the next 64 bits).
	Given $k$, the challenge is then to solve for a suitable 32-bit nonce of $M$ that results in a hash $H$ with the first $k$ bits being 0. We construct challenges in this manner because Bitcoin uses 32-bit nonces to solve for hashes starting with varying $k$ zeroes.
        See Fig.~\ref{fig:bitcoin} for an illustration.
	We generate instances for $k = \{10,15,20\}$ using the generic ANF encoding available at \url{https://github.com/vsklad/cgen}.
%}	
	
		\subsection{Instances from SAT 2017 Competition}
		
%		We selected 25 CNF instances from main track of the SAT 2017 competition benchmarks \cite{balyo2017sat}.\footnote{\url{https://baldur.iti.kit.edu/sat-competition-2017/benchmarks/Main.zip}.} In particular, we omitted instances with larger than 10,000 clauses due to the long conversion time from CNF to the internal ANF representation in \textsc{PolyBoRi}. The primary purpose of our usage of SAT instances is to show that the architecture of iterative bridge between ANF and CNF solving has potential for impact beyond the standard ANF instances. 
                We preprocess \texttt{\small g2-hwmcc15deep-beemfwt4b1-k48} and \texttt{\small g2-hwmcc15deep-beemlifts3b1-k29} using CryptoMiniSat5 to reduce the number of variables to less than 1,048,574 variables,
                which is the maximum number of variables that the \textsc{PolyBoRi} data structure can handle on our platforms.
                We omit the 40 CNFs with names of the pattern \texttt{\small g2-T\textasteriskcentered} because they each have too many variables even after the preprocessing.                
		We also omit \texttt{\small mp1-bsat222-777} because it is not a well-formed DIMACS file.
                Hence, we experiment on 310 instances altogether.
                From these, we select difficult instances: using the runtime of MiniSat (without {\Bosphorus}) as a proxy difficulty measure, we select the 219 that requires more than 2,500 seconds.

\end{document}